\documentclass[lettersize,journal]{IEEEtran}
\usepackage{amsmath,amsfonts}
\usepackage{algorithmic}
\usepackage{algorithm}
\usepackage{array}
\usepackage[caption=false,font=normalsize,labelfont=sf,textfont=sf]{subfig}
\usepackage{textcomp}
\usepackage{stfloats}
\usepackage{url}
\usepackage{verbatim}
\usepackage{graphicx}
\usepackage{cite}
\usepackage{tabularx} 
\usepackage{booktabs}
\usepackage{amssymb} 
\usepackage{booktabs} 
\usepackage{balance}
\begin{document}

\title{Beyond Universality: The GCC-FER Dataset and Culture-Aware Adaptation for Dynamic Facial Expression Recognition}
\author{
Sonalika Singh,
Jyotirindra Dandapat,
Avishi Razdan,
Kshipra V. Moghe,
Puneet Gupta,
and Lalan Kumar
\thanks{
Sonalika Singh, Jyotirindra Dandapat, and Lalan Kumar are with the Department of Electrical Engineering, Indian Institute of Technology Delhi, India.
}
\thanks{
Puneet Gupta is with the Department of Computer Science and Engineering, Indian Institute of Technology Indore, India.
}
\thanks{
Avishi Razdan and Kshipra V. Moghe are with the Department of Psychology, COEP Technological University, India.
}
}

\maketitle

\begin{abstract}
Dynamic Facial Expression Recognition (DFER) is a key enabling technology in affective computing, human-computer interaction, and intelligent multimedia systems. Despite the significant influence of cultural nuances on FER performance, most existing static/dynamic FER systems assume that emotional expressions are universally consistent across populations. The cross-cultural FER performance variation can be attributed to systematic variability in facial muscle activation patterns across cultures. A major challenge in advancing cross-cultural FER lies in the scarcity of culturally diverse benchmark datasets. To address this limitation, a new hybrid multicultural video dataset, termed Global Cross-Cultural Facial Expression Recognition (GCC-FER), is introduced and will be made publicly available to the research community. GCC-FER comprises 23,934 video samples spanning four cultural groups (African, Caucasian, East Asian, and South Asian) across seven basic expressions. The dataset is constructed using a hybrid curation strategy that combines psychologically supervised in-house data collection for historically underrepresented African and South Asian populations alongside rigorous ethnicity filtering of existing video sources. To the best of our knowledge, GCC-FER is the first large-scale global cross-cultural DFER dataset specifically designed to address these demographic gaps. Leveraging this diverse dataset, behaviorally grounded cultural priors are derived separately for each cultural group along with a global prior for practical deployment scenarios. Based on these priors, a Culture-Aware FER (CA-FER) system is proposed to mitigate cultural bias by adaptively recalibrating latent facial representations. Extensive experiments on GCC-FER and DFEW datasets demonstrate that GCC-FER provides a challenging benchmark for cross-cultural FER, while the proposed CA-FER system consistently improves FER performance across multicultural settings. The GCC-FER dataset and associated code will be made publicly available upon paper acceptance at 
\url{https://github.com/SonalikaSingh/GCCFER}.
\end{abstract}


\section{Introduction}
Facial expression recognition (FER) from videos has become an essential component of modern multimedia systems, enabling applications such as affective computing, intelligent human–computer interaction, behavioral monitoring, and multimedia content analysis \cite{corneanu2016survey}. Recent advances in deep learning and large-scale video datasets have significantly improved FER performance by enabling models to learn spatio-temporal representations from facial dynamics \cite{corneanu2016survey, scientificreports2024}. In particular, transformer-based architectures and video-based deep learning models have demonstrated strong capabilities in capturing temporal facial cues across consecutive frames \cite{islam2025facial}. Despite these advances, most existing FER approaches implicitly assume that emotional expressions are universally expressed across different populations \cite{ekman1992argument}. However, psychological and behavioral studies suggest that facial expressions may vary across cultures due to differences in social norms, display rules, and facial muscle activation patterns \cite{matsumoto2008culture}. Ignoring such cross-cultural variations can limit the robustness and generalization of FER systems when applied to diverse multimedia data from the real-world. Moreover, most existing FER systems assume cultural universality or rely on explicit cultural labels, which may not be available in real-world deployments. This limits their applicability in diverse and unconstrained environments.

Early psychological studies suggested that basic emotions may be universally expressed and recognized across cultures \cite{ekman1971constants}. However, subsequent cross-cultural investigations have reported systematic variability in both the production and perception of facial expressions \cite{elfenbein2002above, Facial_expressions_of_emotion}. The rules of cultural display influence how emotions are expressed, regulated, and interpreted across societies \cite{matsumoto1992american}. Furthermore, experimental studies have shown that observers from different cultural backgrounds rely on different facial regions when interpreting emotional expressions \cite{Jack2012}. These findings suggest that emotional expression is shaped not only by biological mechanisms but also by socio-cultural norms and perceptual strategies. Despite extensive evidence of cultural variability in psychology, its systematic incorporation into computational FER frameworks remains limited.

Dataset composition is key to the performance of deep learning-based FER systems. Widely used datasets such as CK+ \cite{lucey2010extended}, real-world affective faces database (RAF-DB) \cite{raf-db}, AffectNet \cite{mollahosseini2017affectnet}, acted facial expressions in the wild (AFEW) \cite{kossaifi2017afew}, and dynamic facial expression in the wild (DFEW) \cite{jiang2020dfew} have significantly advanced the field by providing annotated facial expressions collected under both controlled and in-the-wild conditions. Recent cross-cultural studies further reveal that FER varies significantly across populations, with notable differences observed between Western groups (e.g. Germans, Australians) and underrepresented or less-exposed populations such as the Karamojong and Malaysian population \cite{krippl2026, mohan2021exploratory}. Complementary efforts to introduce culturally specific datasets, such as Indian micro-expression data \cite{mishra2025toward}, remain limited in scale and diversity. However, many of these datasets are disproportionately composed of Western and East Asian subjects \cite{ggarcia2022macro}. Such biases can limit the generalization capability of FER models when applied to underrepresented populations. Consequently, performance improvements observed on benchmark datasets do not always translate to robust cross-cultural deployment.

Several studies have attempted to address cultural variability in FER systems. Approaches including multicultural annotation strategies, dynamic action unit (AU) modeling, and label distribution learning have been explored to account for population-level variation. Cross-cultural evaluation studies have further demonstrated measurable performance degradation when models trained on one demographic group are tested on another \cite{bonassi2021cross}. Recent analyses further reveal systematic cultural gaps in feature-based FER systems and highlight cross-race inconsistencies in model predictions \cite{siritanawan2023exploring, li2023cross}. This further, indicates that learned representations are not invariant across populations . While these studies consistently establish that FER performance is influenced by cultural factors, most existing approaches primarily focus on analyzing or mitigating bias at the data or evaluation level, without explicitly incorporating cultural context into the model design. As a result, temporal expression dynamics and region-specific cues may not be adequately captured across diverse populations.

Deep learning has significantly improved FER performance in recent years. In particular, convolutional neural networks (CNNs) \cite{wu2017introduction} are widely used for this purpose because of its ability to extract the spatial features from facial images and video frames. FER-YOLO \cite{jin2026fer} demonstrates the effectiveness of CNN-based architectures for real-time FER system.  Additionally, 3D CNNs \cite{3dcnn} and hybrid architectures have been introduced to capture temporal information. Vision transformers have recently emerged as an effective alternative for video-based recognition tasks. Architectures such as video vision transformer (ViViT) \cite{arnab2021vivit} and efficient transformer variants \cite{liu2023efficientvit, cheng2023hla} leverage self-attention mechanisms to model global spatio-temporal relationships. Recent FER research has explored the incorporation of auxiliary information and advanced learning paradigms to improve expression representation learning. CEPrompt \cite{zhou2024ceprompt} integrates emotion conception derived from vision-language pretraining into facial appearance representations through cross-modal prompting, enabling enhanced discrimination of subtle facial expressions. In a different direction, AMY \cite{ding2024uncertainty} addresses personalized federated facial expression recognition by modeling uncertainty and high-order relationships among expression samples using hypergraphs, thereby improving robustness under decentralized and heterogeneous data distributions. However, these methods do not explicitly account for cultural variability in facial expression perception and representation. Underrepresented FER data sets, along with an absence of cultural adaptation, make model generalization very limited with performance degradation. 

These limitations suggest the need for a comprehensive cross-cultural dataset and a system that jointly addresses spatio-temporal modeling and cultural adaptation in FER. Motivated by these challenges, this work presents a novel dataset and proposes a culturally aware spatio-temporal FER system that integrates cross-cultural facial expression into video-based expression recognition.
\noindent{The main contributions of this work are summarized as follows:}

\begin{itemize}

\item A large-scale multicultural dynamic facial expression recognition dataset, termed GCC-FER, is introduced comprising 23,934 video samples across four major cultural groups and seven facial expression categories. To the best of our knowledge, GCC-FER is the first large-scale global DFER benchmark specifically designed for systematic cross-cultural facial expression analysis.

\item A Culture-Aware Facial Expression Recognition (CA-FER) framework is proposed that leverages behaviorally grounded cultural priors to mitigate cultural bias in dynamic facial expression recognition.

\item Extensive cross-cultural experiments on GCC-FER and DFEW establish behaviorally grounded cultural priors as an effective mechanism for mitigating cultural bias in dynamic facial expression recognition.

\end{itemize}
\section{Dataset}
This section introduces GCC-FER, a multicultural video dataset designed to support cross-cultural facial expression recognition and the analysis of cultural variability in facial behavior \cite{ICC2025}. To contextualize the contribution of GCC-FER, Table~\ref{tab:dataset_comparison} compares it with representative FER and DFER benchmarks. While recent large-scale video datasets such as DFEW, MAFW, and FERV39K have significantly advanced dynamic facial expression recognition, they do not provide explicit cultural annotations and are not designed for cross-cultural analysis. Conversely, ICC-FER investigates cultural variability within the Indian context by considering regional cultural groups; however, it focuses on frame-level image analysis rather than temporal expression dynamics and does not address global cross-cultural variation. GCC-FER bridges this gap by providing a large-scale multicultural video benchmark comprising 23,934 video samples across African, Caucasian, East Asian, and South Asian populations, thereby enabling the study of cross-cultural dynamic facial expression recognition at a global scale.
\begin{table}[t]
\caption{Comparison of representative FER datasets.}
\label{tab:dataset_comparison}
\small
\setlength{\tabcolsep}{1pt}
\centering
\begin{tabular}{|l|c|c|c|c|}
\hline
Dataset & Samples & Dynamic & Explicit Culture & Cross-Cultural\\
\hline
CK+  \cite{lucey2010extended}      & 920      & $\times$ & $\times$ & $\times$ \\
RAF-DB \cite{raf-db}    & 29,672   & $\times$ & $\times$ & $\times$ \\
AffectNet \cite{mollahosseini2017affectnet} & 1.5M   & $\times$ & $\times$ & $\times$ \\
AFEW  \cite{kossaifi2017afew}     & 1,809    & $\checkmark$ & $\times$ & $\times$ \\
DFEW  \cite{jiang2020dfew}    & 16,372   & $\checkmark$ & $\times$ & $\times$ \\
MAFW   \cite{mafw}    & 10,045   & $\checkmark$ & $\times$ & $\times$ \\
FERV39K \cite{ferv39k}  & 38,935   & $\checkmark$ & $\times$ & $\times$ \\
ICC-FER \cite{ICC2025}   & 70,000    & $\times$ & $\checkmark$ & $\checkmark$ \\
{GCC-FER}$^{a}$ & \textbf{23,934} &
\textbf{$\checkmark$} &
\textbf{$\checkmark$} &
\textbf{$\checkmark$} \\
\hline
\end{tabular}

\vspace{1mm}
\raggedright
\footnotesize
$^{a}$ Proposed dataset.
\end{table}
\subsection{Collection Protocol}
The GCC-FER dataset spans four major cultural groups: African, Caucasian, East Asian, and South Asian, comprising a total of 23,934 video samples. The South Asian and African subsets, consisting of 13,577 video clips, were curated in-house from publicly available YouTube videos released under Creative Commons licenses. The East Asian and Caucasian subsets, comprising 10,357 samples, were adopted from the DFEW dataset with additional ethnicity filtering and verification following approval from the authors.
Each sample consists of a five-second video segment recorded at 720p resolution. Video clips containing more than two individuals were excluded to reduce ambiguity in facial expression analysis \cite{liu2023facial}. For segments containing up to two persons, the individual appearing most frequently in the clip was selected for further processing. Face detection and alignment were performed using MediaPipe \cite{lugaresi2019mediapipe}. Uniform temporal sampling \cite{scientificreports2024} was applied to extract 16 face frames per video. Each frame was resized to $224 \times 224$ pixels and normalized using ImageNet statistics before being used as input to the ViViT model.
\subsection{Annotation Protocol}
\subsubsection{Ethnicity Filtering}
Ethnic consistency in the dataset was ensured using a multi-stage filtering protocol. Two trained annotators independently labeled each video according to four cultural categories: Caucasian, East Asian, South Asian, and African. The annotations achieved 
substantial agreement with Cohen's $\kappa = 0.786$ and 89.4\% overall agreement. Automated ethnicity estimation was then performed using the DeepFace framework. A video sample was retained only if at least two of the three sources 
(two annotators and DeepFace with confidence $\ge 75\%$) agreed on the cultural label. To further validate the reliability of the filtering process, a ResNet-18 classifier was 
trained on the filtered features to evaluate ethnic separability. The classifier achieved 88.0\% accuracy using 5-fold cross-validation, confirming consistent cultural grouping.
\subsubsection{Expression Annotation}
Facial expressions were annotated by four trained annotators under the supervision of an expert psychologist. The annotation protocol integrates Ekman's Basic Emotion Theory \cite{ekman1993facial} with the Feeling Wheel and Geneva Emotion Wheel frameworks. Each video clip was labeled as one of seven basic emotions according to the dominant expression observed in the segment. Clips with annotation disagreement were excluded to ensure label reliability \cite{jiang2020dfew}. Inter-annotator agreement was evaluated using Fleiss' Kappa coefficient. The South Asian subset achieved Kappa values between 0.6966 and 0.7037 (average $0.7010 \pm 0.0031$), while the African subset showed agreement between 0.6329 and 0.6443 (average $0.6372 \pm 0.0051$). According to the Landis–Koch guidelines \cite{landis1977measurement}, these values indicate substantial agreement, validating the annotation reliability.
\begin{table*}{}
\caption{Distribution of the GCC-FER dataset across cultural groups and expression categories.}
\label{tab:dataset_distribution}
\centering
\small
\begin{tabular}{|l|c|c|c|c|c|c|c|c|}
\hline
\textit{Culture} & \textit{Angry} & \textit{Disgust} & \textit{Fear} & \textit{Happy} & \textit{Neutral} & \textit{Sad} & \textit{Surprise} & \textit{Total} \\
\hline
Caucasian   & 1,041 & 299  & 597  & 1,206 & 986   & 881  & 791  & 5,801 (24.2\%) \\

East Asian  & 747   & 257  & 481  & 890   & 630   & 908  & 643  & 4,556 (19.0\%) \\

South Asian & 844   & 285  & 303  & 1,288 & 2,326 & 1,109 & 471 & 6,626 (27.7\%) \\

African     & 828   & 437  & 255  & 2,040 & 1,977 & 1,163 & 251 & 6,951 (29\%) \\

\hline
\textit{Total} & 3,460 & 1,278 & 1,636 & 5,424 & 5,919 & 4,061 & 2,156 & 23,934 \\
\hline

\end{tabular}
\end{table*}

\subsection{Dataset Statistics}

The GCC-FER dataset contains 23,934 video samples distributed across four cultural groups and seven expression categories. Table~\ref{tab:dataset_distribution} presents the 
distribution of samples across both cultural groups and expression classes.

Neutral expressions constitute the largest portion of the dataset (24.7\%), followed 
by Happy (22.7\%), while Disgust (5.3\%) and Fear (6.8\%) are less frequent. 
This distribution reflects natural expression frequencies commonly observed in real-world social interactions. The dataset exhibits moderate class imbalance, particularly for minority expression categories. To mitigate potential bias during training, focal loss is employed to emphasize harder samples and improve discrimination for underrepresented classes. Furthermore, stratified 5-fold cross-validation is adopted to preserve the 
distribution of expression–culture combinations across training and validation splits.
\subsection{Cultural Representation}

Cultural representation in the dataset is relatively balanced, with African samples 
accounting for 29\%, South Asian 27.7\%, Caucasian 24.2\%, and East Asian 19.0\%  of the total dataset.Although the distribution is reasonably balanced, East Asian samples remain 
comparatively underrepresented, reflecting a common bias observed in many large-scale affective computing datasets. 
The use of stratified cross-validation helps ensure that all culture–expression combinations are proportionally represented during model evaluation.

\begin{figure}
\centering
\includegraphics[width=1\linewidth]{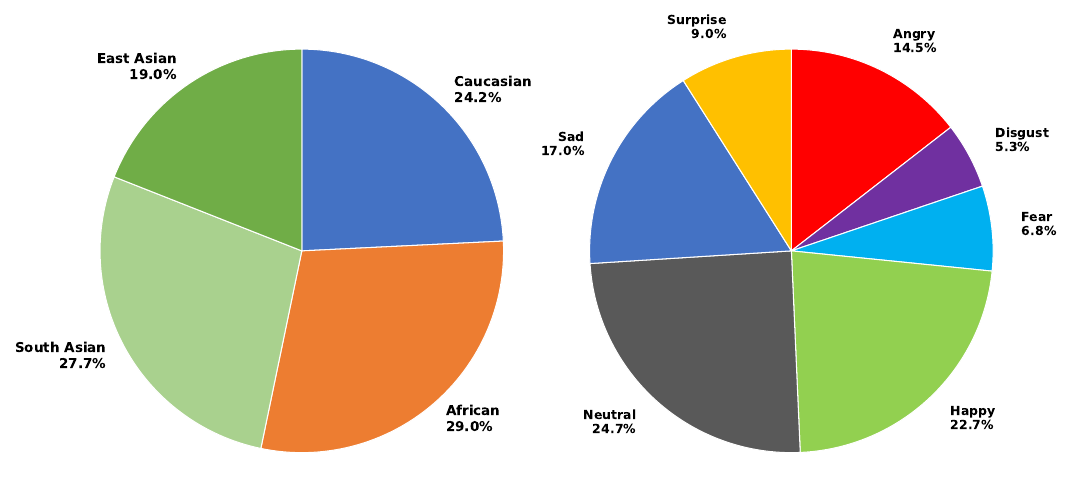}
\caption{Proportional distribution of the GCC-FER dataset across cultural groups (left) and expression categories (right).}
\label{fig:gcc_distribution}
\end{figure}

To provide a clearer visualization of the dataset composition, Fig.~\ref{fig:gcc_distribution} illustrates the proportional distribution of samples across cultural groups and expression categories in the GCC-FER dataset.

\subsection{Ethical statement}
The South Asian and African ethnicity part of the data in the GCC-FER dataset was constructed exclusively from publicly available YouTube videos released under Creative Commons licenses. No direct interaction with human participants was involved. The study, therefore, relies solely on a retrospective secondary analysis of publicly available multimedia content. To protect privacy, no personal identifiable information (PII) such as names, geolocation metadata, or user identities were collected. All video samples were assigned anonymous identifiers during dataset construction. The dataset and models are intended solely for research on FER and cross-cultural affective computing. No identity recognition, biometric verification, or individual profiling is performed. If any copyright holder believes that their rights have been infringed, the corresponding material will be promptly reviewed and removed from the dataset upon request. It should be noted that the East Asian and Caucasian data in the GCC-FER dataset was adopted from the DFEW dataset. The missing ethnicity analysis was conducted for this dataset at our end, post-appropriate permission.    

An application for ethical review has been submitted to the Institute Ethics Committee (IEC) at IIT Delhi, and the review process is currently underway. All data collection and usage procedures follow institutional guidelines for secondary analysis of publicly available human-subject data.

\section{System Model}
\begin{figure*}[!t]
    \centering
    \includegraphics[
        width=\textwidth,
        trim=0cm 0cm 0cm 1cm,
        clip
    ]{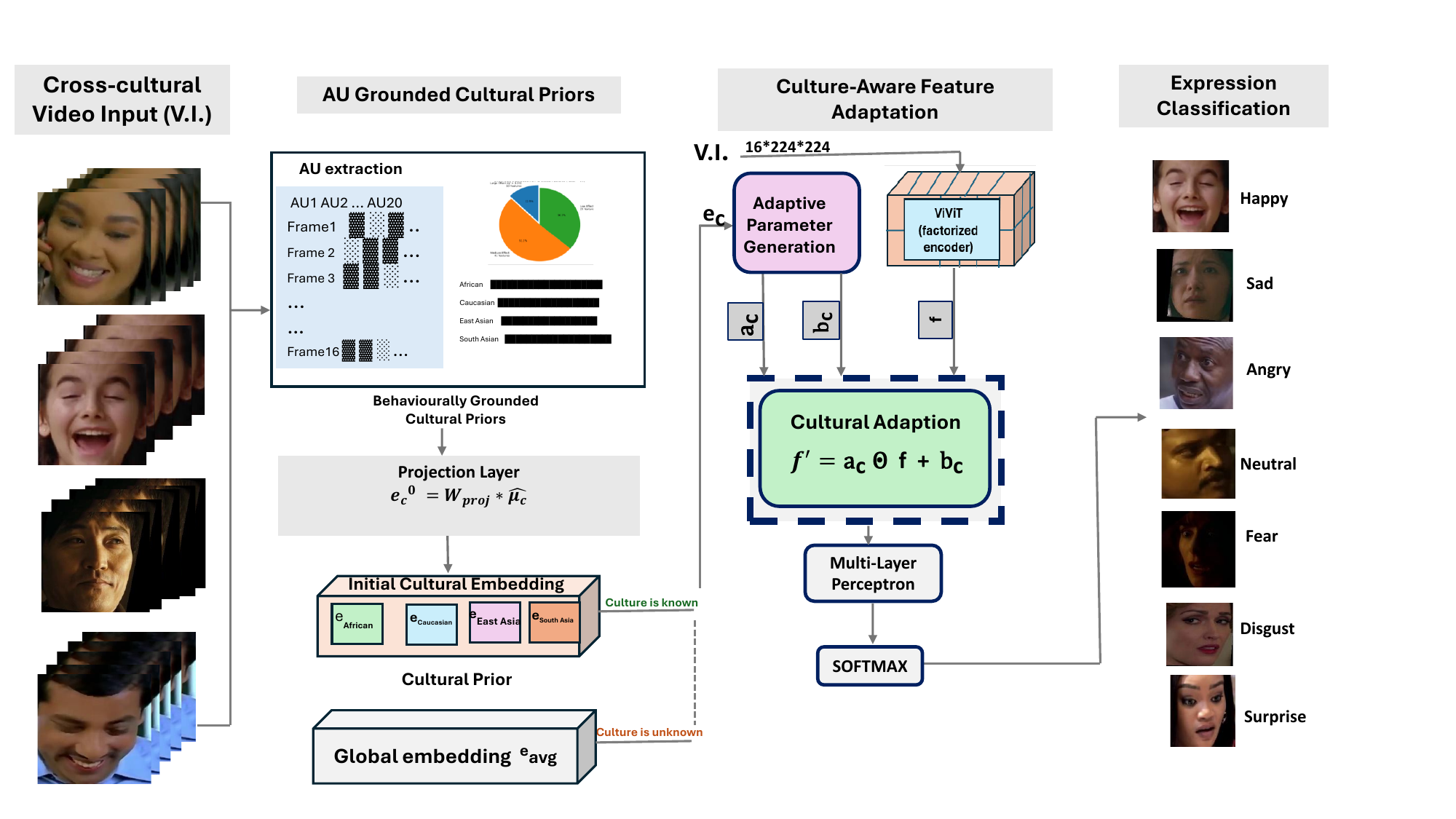}
    \caption{Architecture of the proposed CA-FER system for culture-aware dynamic facial expression recognition. Cross-cultural facial behavior analysis is utilized to construct behaviorally grounded cultural priors, which adapt latent spatio-temporal representations extracted by the ViViT backbone for mitigating cross-cultural representation bias.}
    \label{fig:architecture}
\end{figure*}

This section describes the proposed CA-FER system for culture-aware dynamic FER. The proposed system mitigates cross-cultural representation bias in FER by incorporating behaviorally grounded cultural priors into a spatio-temporal recognition model. The overall system consists of two main stages: (i) behaviorally grounded cultural prior modeling using facial Action Unit (AU) activation patterns to derive culture-specific latent priors, and (ii) adaptation of latent facial representations within the learned feature space using these behavioral priors. An overview of the proposed system is illustrated in Figure~\ref{fig:architecture}.

\subsection{System Overview}
Given an input video sequence $\mathit{X}$ containing $T$ facial frames, the goal is to predict the corresponding expression state $y$ by leveraging behaviorally grounded cultural priors. As described in Section III, the system operates in two stages. In the first stage, AU intensity analysis is performed to derive culture-specific behavioral priors along with a global behavioral prior across all cultural groups. In the second stage, these behavioral priors are incorporated to adapt latent spatio-temporal representations extracted by the ViViT backbone. This enables the proposed system to mitigate cross-cultural representation bias during facial expression recognition compared to conventional culture-agnostic systems.

\subsection{Behaviorally Grounded Cultural Prior Modeling}
AUs provide an interpretable representation of facial muscle movements underlying emotional expressions. For each video sample $\mathit{X}_i$, frame-level AUs are extracted using the PyFeat framework \cite{cheong2023py}. Given uniformly $T$ sampled frames per video, the AU activation matrix is represented as

\begin{equation}
\mathit{V}_i \in \mathbb{R}^{T \times N_a},
\end{equation}
where $N_a$ denotes the number of Action Units (AUs) considered in the analysis (20 in this work).
To summarize temporal AU behavior within a video, four statistical measures are
computed for each AU across the frame sequence: mean activation, standard
deviation, maximum activation, and activation frequency. These statistics are
concatenated to form an 80-dimensional feature vector as

\begin{equation}
\mathit{v}_i = [\mu, \sigma, \max, f]_{\text{AU}} \in \mathbb{R}^{80}.
\end{equation}
To quantify cultural variability in facial behavior, one-way ANOVA test is performed across cultural groups for each AU statistic. The resulting $F$-statistics, p-values, and effect sizes ($\eta^2$) measure the degree to which facial muscle activation patterns differ across cultures. For culture c, the AU profile is defined as the mean AU activation across all samples belonging to that culture, based on empirical observations.
\begin{equation}
\boldsymbol{\mu}_c =
\frac{1}{N_c} \sum_{i:c_i=c} \mathit{v}_i^{\text{mean}},
\end{equation}
Here, $N_c$ denotes the number of samples in culture $c$ and
$\mathit{v}_i^{\text{mean}}$ represents the mean AU activation vector across the frames.
To ensure consistent scaling across cultures, the cultural AU profile is
normalized using min--max normalization as
\begin{equation}
\hat{\boldsymbol{\mu}}_c =
\frac{\boldsymbol{\mu}_c - \boldsymbol{\mu}_{\min}}
{\boldsymbol{\mu}_{\max} - \boldsymbol{\mu}_{\min}}.
\end{equation}
The normalized AU profile is then projected into a $d_e$-dimensional embedding
space via a linear transformation, where $d_e$ denotes the cultural embedding dimension
($d_e=128$ in this work)
\begin{equation}
\mathbf{e}_c^{(0)} = W_{\text{proj}} \hat{\boldsymbol{\mu}}_c^{T},
\end{equation}
where $W_{\text{proj}} \in \mathbb{R}^{d_e \times 20}$.
Here, $\mathbf{e}_c^{(0)}$ denotes the initial cultural embedding for culture $c$, which is initialized from the corresponding AU activation profile. This serves as the starting representation for culture $c$ and is subsequently updated during end-to-end training via backpropagation, yielding the learned cultural embedding $\mathbf{e}_c$. The learned embedding is then utilized to generate culture-specific transformation adaptation parameters:
\begin{equation}
a_c = W_a \mathbf{e}_c, \quad
b_c = W_b \mathbf{e}_c,
\end{equation}
where $W_a, W_b \in \mathbb{R}^{768 \times d_e}$ are learnable parameter matrices.

\subsection{Culture-Agnostic Baseline System}
The Culture-Agnostic baseline system employs the ViViT Model-3 (Factorised Encoder) architecture \cite{arnab2021vivit} for dynamic facial expression recognition without explicit cultural adaptation. The network employs tubelet embedding with temporal stride 2 and spatial patch size $16 \times 16$ at an input resolution of $224 \times 224$. Given an input video sequence:
\begin{equation}
\mathit{X} \in \mathbb{R}^{T \times H \times W \times 3},
\end{equation}
the ViViT backbone extracts latent spatio-temporal token representations from the final transformer layer as

\begin{equation}
F = \{f_{\text{CLS}}, f_1, f_2, \dots, f_N\},
\end{equation}
where $f_{\text{CLS}} \in \mathbb{R}^{768}$ denotes the global classification token and $\{f_i\}_{i=1}^{N}$ represent spatial patch token embeddings. The spatial patch tokens are aggregated using average pooling:
\begin{equation}
f_{\text{avg}} = \frac{1}{N}\sum_{i=1}^{N} f_i.
\end{equation}
To jointly encode global and local spatio-temporal information, the CLS token and aggregated patch representation are concatenated and projected into a unified latent representation:
\begin{equation}
f = W_f [f_{\text{CLS}} ; f_{\text{avg}}] + b_f,
\end{equation}
where $[\cdot ; \cdot]$ denotes feature concatenation. The resulting latent representation $f$ is subsequently passed through a two-layer multilayer perceptron (MLP) with ReLU activation and dropout ($p=0.3$) for expression classification. Finally, the predicted probability distribution over the seven expression categories is obtained using the softmax function:
\begin{equation}
\hat{y} = \text{Softmax}(W_c f + b_c).
\end{equation}

\subsection{Culture-Aware Representation Adaptation and Expression Classification}
While the culture-agnostic baseline extracts latent representation without cultural awareness, the proposed CA-FER system recalibrates these features using behaviorally grounded cultural priors derived from AU activation profiles.
The latent spatio-temporal representation extracted by the ViViT backbone is adaptively recalibrated using behaviorally grounded cultural priors. This further mitigates cross-cultural representation bias during facial expression recognition. Unlike conventional approaches that initialize embeddings randomly \cite{perez2018film}, the proposed system initializes cultural embeddings from cross-cultural facial muscle activation patterns. This enables the latent feature space to incorporate population-level facial behavioral priors during representation learning. The latent representation $\mathbf{f}$ is then adaptively recalibrated as  
\begin{equation}
\mathbf{f'} = a_c \odot \mathbf{f} + b_c,
\end{equation}
where $\odot$ denotes element-wise multiplication.
This adaptive transformation enables the learned feature space to incorporate culturally observed facial behavioral characteristics during spatio-temporal facial representation learning. The recalibrated feature vector $\mathit{f}'$ is passed through a two-layer classifier to predict the final expression category as

\begin{align}
\mathit{h} &= \text{ReLU}(W_1 \mathit{f}' + b_1),\\
\mathit{z} &= W_2 \mathit{h} + b_2\\
P(y|\mathit{X},c) &= \text{softmax}(\mathit{z})
\end{align}

where $y \in \{1,2,\ldots,7\}$ denotes the predicted facial expression label among the seven expression classes.

To address class imbalance in expression datasets, focal loss is used during training. During training, all model parameters, including cultural embeddings and transformer weights, are optimized using gradient-based learning. The complete training pipeline of the proposed CA-FER system is summarized in algorithm~\ref{alg:cafer}. 

\begin{algorithm}[t]
\caption{CA-FER: Behaviorally Grounded Cultural Prior Adaptation for Culture-Aware Facial Expression Recognition}
\label{alg:cafer}
\begin{algorithmic}[1]

\REQUIRE Dataset $\mathcal{D}=\{\mathit{X}_i,c_i,y_i\}$
\ENSURE Expression prediction $\hat{y}$

\STATE \textbf{Phase 1: Behaviorally Grounded Cultural Prior Modeling (Offline)}

\FOR{$i = 1$ to $N$}
    \STATE $\mathit{V}_i \leftarrow \text{PyFeat}(\mathit{X}_i)$
    \STATE $\mathit{v}_i \leftarrow [\mu,\sigma,\max,f]_{\text{AU}}$
\ENDFOR

\STATE $(F_k,p_k,\eta_k^2) \leftarrow 
\text{ANOVA}(\{\mathit{v}_i\},\{c_i\})$

\FOR{each culture $c$}
    \STATE $\mu_c \leftarrow \frac{1}{N_c}
    \sum_{i:c_i=c}\mathit{v}_i^{mean}$

    \STATE $\hat{\mu}_c \leftarrow 
    \text{Normalize}(\mu_c)$

    \STATE $\mathbf{e}_c^{(0)} \leftarrow 
    W_{proj}\hat{\mu}_c$
\ENDFOR

\STATE Initialize trainable embedding table:
\STATE $E \leftarrow \{\mathbf{e}_c^{(0)}\}$

\vspace{0.5mm}
\STATE \textbf{Phase 2: Culture-Aware Latent Representation Adaptation}

\FOR{each mini-batch $\mathcal{B}$}

    \STATE $\mathbf{f} \leftarrow 
    \text{ViViT}(\mathit{X})$

    \STATE Retrieve refined cultural embedding:
    \STATE $\mathbf{e}_c \leftarrow E[c]$

    \STATE Generate scaling and shifting parameters:
    \STATE $a_c \leftarrow W_a \mathbf{e}_c$
    \STATE $b_c \leftarrow W_b \mathbf{e}_c$

    \STATE Adapt latent representation:
    \STATE $\mathbf{f}' \leftarrow 
    a_c \odot \mathbf{f} + b_c$

    \STATE $\hat{y} \leftarrow 
    \text{Softmax}(\text{Classifier}(\mathbf{f}'))$

    \STATE Update ViViT, embedding table $E$, classifier,
    $W_a$, and $W_b$ using focal loss

\ENDFOR

\RETURN $\hat{y}$

\end{algorithmic}
\end{algorithm}

\subsection{Cultural Adaptation without Explicit Cultural Labels}
In practical deployment scenarios, explicit cultural labels may not always be available or may contain noise. To address this limitation, the proposed system utilizes a global AU-grounded cultural prior derived from the mean facial Action Unit activation profile aggregated across all cultural groups. This global behavioral prior serves as a generalized initialization for latent representation adaptation in the absence of explicit cultural supervision. Consequently, the system retains the benefits of behaviorally grounded cultural adaptation while operating under culture-agnostic inference settings. As demonstrated in Figure~\ref{fig:granularity}, the proposed strategy consistently improves performance over the conventional culture-agnostic baseline, highlighting the robustness and practical applicability of behaviorally grounded cultural priors for cross-cultural FER.
\section{Experimental Conditions}\label{sec:exp}
All experiments were conducted on a workstation equipped with four NVIDIA GeForce RTX 2080 Ti GPUs (11GB VRAM each). The system utilized CUDA 12.4 with driver version 550.107.02 for optimal deep learning performance. The proposed CAFER system was trained using the AdamW optimizer with a learning rate of $10^{-5}$, a batch size of 8 (distributed across 4 GPUs), and early stopping with a patience of 7 epochs. Label smoothing (0.1) and gradient clipping (max norm 1.0) were applied for stable training. Focal loss with label smoothing ($\varepsilon = 0.1$) was utilised as the training objective to jointly address class imbalance and overconfident predictions. Focal loss dynamically down-weights well-classified samples and focuses learning on harder, underrepresented expression classes. Label smoothing further prevents overconfidence by creating soft targets. This enhances generalisation across cultural variations in emotional expressions.
\section{Results and Discussion}
\label{sec:results}
This section presents a comprehensive evaluation of the proposed CA-FER system on the in-house collected GCC-FER dataset and the publicly available DFEW benchmark. In particular, effectiveness of AU-grounded cultural embedding initialization is studied and compared with state-of-the-art methods. Unweighted Average Recall (UAR) is reported as the primary metric to handle class imbalance, and Weighted Average Recall (WAR) \cite{jiang2020dfew} is reported for overall accuracy assessment~\cite{mma}. UAR is emphasized as it assigns equal importance to all
expression categories irrespective of class frequency. Ablation studies are additionally presented to validate key design choices.

\subsection{Evaluation on the GCC-FER Dataset}
Table~\ref{tab:gccfer_results} presents the comprehensive evaluation on the GCC-FER dataset under 5-fold cross-validation. Among the culture-agnostic methods, DPCNet achieves the best performance with 56.12\% UAR and 59.47\% WAR, while the proposed Culture-Agnostic ViViT baseline achieves 54.20\% UAR and 58.10\% WAR. Introducing cultural adaptation consistently improves FER performance across all conditioning strategies. In particular, random embedding initialization improves UAR from 54.20\% to 57.00\%, while one-hot cultural conditioning and separate cultural heads achieve 58.99\% and 59.39\% UAR, respectively.

The proposed CA-FER system achieves the best overall performance with 61.70\% UAR and 64.80\% WAR. Compared with DPCNet (56.12\% UAR), CA-FER improves performance by +5.58 pp in UAR. Compared with the Culture-Agnostic ViViT baseline (54.20\% UAR), the proposed system improves UAR by +7.50 pp. Furthermore, CA-FER outperforms the fixed behavioral prior adaptation strategy (59.54\% UAR) by +2.16 pp, demonstrating the effectiveness of AU-grounded adaptive cultural representation learning for cross-cultural FER.
\begin{table}[t]
\caption{Comprehensive evaluation on GCC-FER dataset (5-fold 
cross-validation). Methods are grouped by conditioning strategy.}
\label{tab:gccfer_results}
\centering
\small
\begin{tabularx}{\columnwidth}{|X|c|c|}
\hline
\textit{Method} & \textit{UAR (\%)} & \textit{WAR (\%)} \\
\hline
\multicolumn{3}{|c|}{\textit{Culture-Agnostic Methods}} \\
\hline
Former-DFER~\cite{former}   & 43.91 & 49.78 \\
NR-DFERNet~\cite{nrfer}     & 44.00 & 51.41 \\
M3DFEL~\cite{M3DFEL}         & 49.73 & 52.81 \\
3D CNN~\cite{3dcnn}          & 54.25 & 56.65 \\
Culture-Agnostic ViViT       & 54.20 & 58.10 \\
DPCNet~\cite{dpcnet}        & 56.12 & 59.47 \\
\hline
\multicolumn{3}{|c|}{\textit{Culture-Adaptation Methods}} \\
\hline
Random Embedding Initialization        & 57.00 & 61.20 \\
One-hot Concatenation~\cite{rodriguez2018beyond} & 58.99 & 63.48 \\
Separate Cultural Heads~\cite{sun2020adashare}   & 59.39 & 64.17 \\
AU-concat Fixed (Proposed)             & 59.54 & 63.80 \\
\textit{CA-FER (Proposed)} & \textit{\textbf{61.70}} & \textit{\textbf{64.80}} \\
\hline
\end{tabularx}
\end{table}
\begin{figure}[t]
\centering
\includegraphics[width=0.45\textwidth, height=0.35\textheight, keepaspectratio]{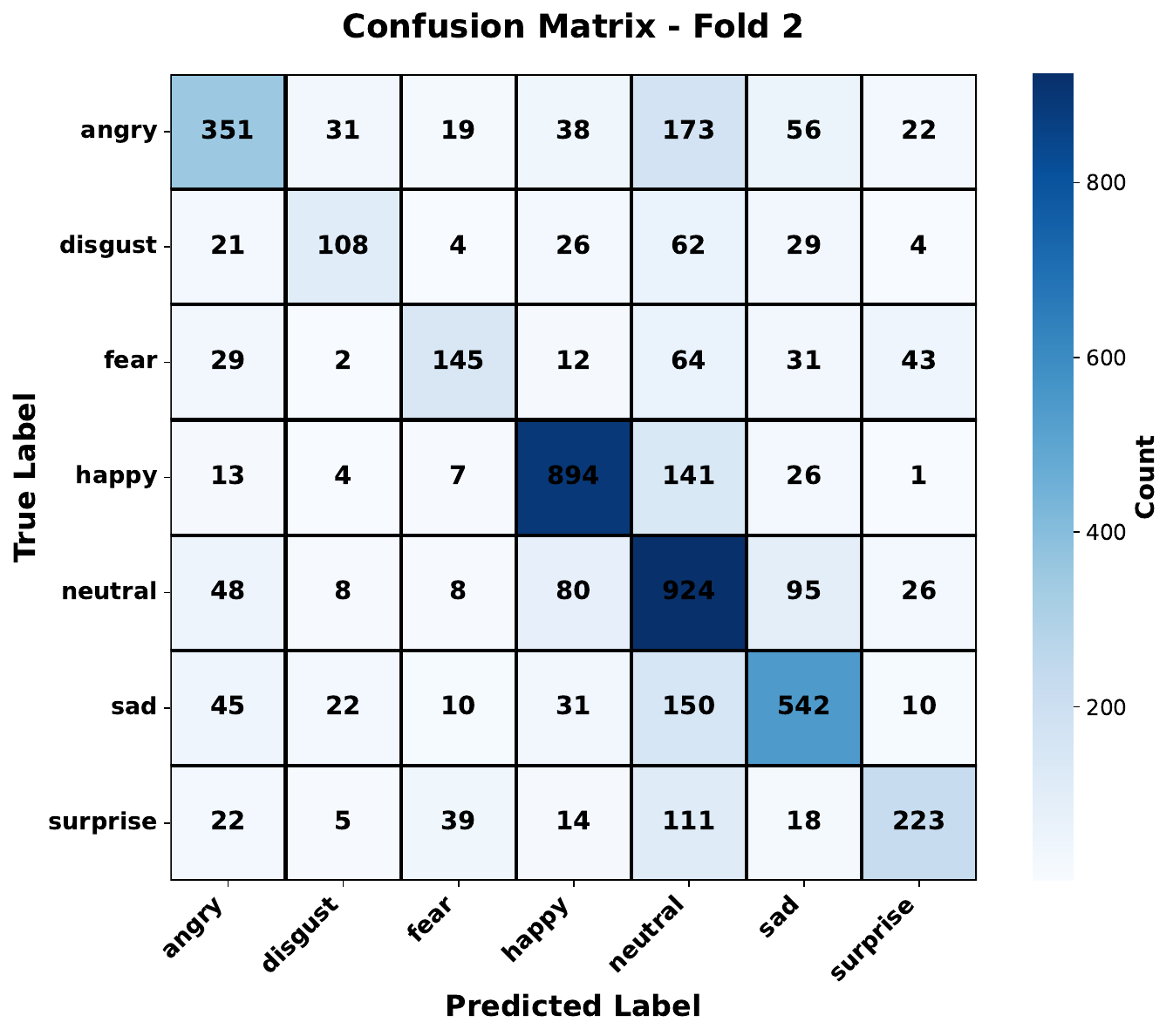}
\caption{Confusion matrix for the GCC-FER dataset (Fold 2 validation). Rows denote ground-truth labels and columns denote predicted classes.}
\label{fig:global_cm}
\end{figure}

The confusion matrix for the GCC-FER dataset is presented in Figure~\ref{fig:global_cm}. The proposed CA-FER system excels at recognizing Happy (82.3\% recall) and Neutral (77.7\%) expressions. Notable confusion occurs between Angry and Neutral (173 instances), reflecting the challenge of distinguishing neutral expressions from expressed anger in naturalistic video data.

\subsection{Benchmark Evaluation on the DFEW Dataset}
\label{ssec:dfew}
To assess cross-dataset generalization, the proposed CA-FER system is evaluated on the publicly available DFEW benchmark. The evaluation follows the standard 5-fold stratified cross-validation protocol used in the dynamic FER (DFER) literature. It should be noted that the state-of-the-art methods rely on large-scale backbone pre-training on datasets such as VoxCeleb2 or ImageNet-21K \cite{nagrani2017voxceleb}. However, the proposed method does not require large-scale external pre-training and yet provides comparable performance. The performance is presented in Table~\ref{tab:dfew_sota}. 

\begin{table}[t]
\caption{Comparison with state-of-the-art visual-only methods on the DFEW dataset (5-fold cross-validation).}
\label{tab:dfew_sota}
\centering
\footnotesize
\setlength{\tabcolsep}{10pt}
\begin{tabular}{|l|c|c|c|}
\hline
\textit{Method} & \textit{UAR} & \textit{WAR} & \textit{Year} \\
\hline

\multicolumn{4}{|c|}{\textit{Methods Using Large-Scale Pre-training}} \\
\hline
S2D~\cite{s2d}              & 65.45 & 76.03 & 2024 \\
MAE-DFER~\cite{mma}         & 63.41 & 74.43 & 2023 \\
SVFAP~\cite{svfap}          & 62.83 & 74.27 & 2024 \\
\hline

\multicolumn{4}{|c|}{\textit{Methods without Large-Scale Pre-training}} \\
\hline
\
DPCNet~\cite{dpcnet}        & 57.11 & 66.32 & 2022 \\
SlowR50-SA~\cite{slow}                  & 57.09 & 69.87  & 2024 \\
M3DFEL~\cite{M3DFEL}        & 56.10 & 69.25 & 2023 \\
IAL~\cite{intensity}        & 55.71 & 69.24 & 2023 \\
NR-DFERNet~\cite{nrfer}     & 55.77 & 68.19 & 2022 \\
STT~\cite{STT}              & 54.58 & 66.65 & 2022 \\
Mamba-style SSM~\cite{mamba}  &54.28 & 66.81 & 2025\\
Former-DFER~\cite{former}   & 53.69 & 65.70 & 2021 \\
\textit{CA-FER (Proposed)}$^{a}$ & \textbf{63.93} & \textbf{65.21} & 2026 \\
\hline
\end{tabular}

\vspace{1mm}
\raggedright
\footnotesize
$^{a}$Best-performing method without large-scale pre-training.
\end{table}
It may be observed that the proposed CA-FER system achieves 63.93\% UAR on DFEW. Among non-pretrained methods, CA-FER outperforms the state-of-the-art dual path collaborative network (DPCNet) \cite{dpcnet} by +6.82 pp in UAR. Additionally, the performance of the proposed CA-FER is comparable with the large-scale pre-trained counterparts with significantly lower data requirements. This shows that behaviorally grounded cultural priors provide stronger representational capacity even without large pretrained backbones. 

\begin{figure}[t]
\centering
\includegraphics[height=0.3\textheight, keepaspectratio]{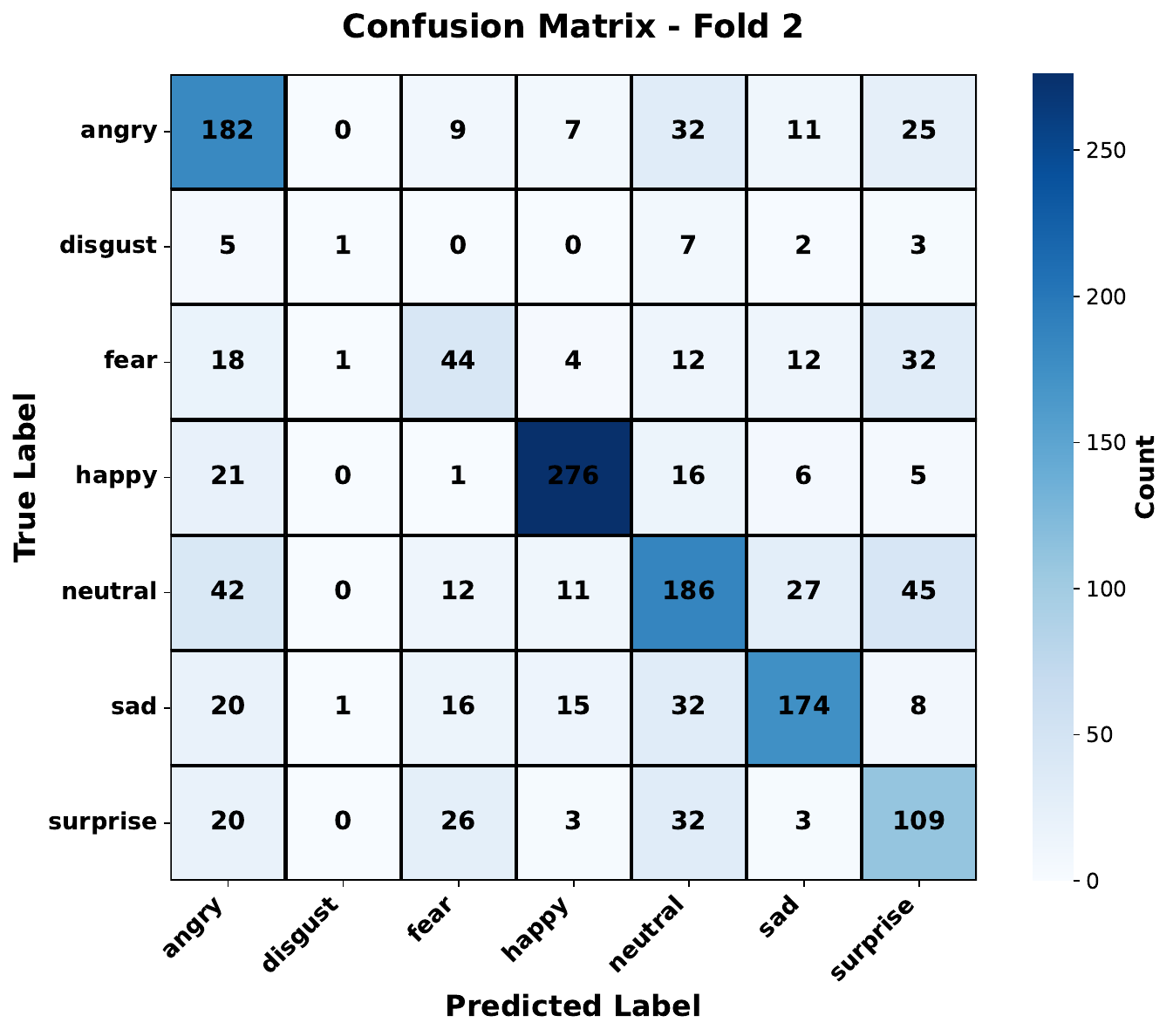}
\caption{Confusion matrix for the DFEW dataset (Fold 2 validation). Rows denote ground-truth labels and columns denote predicted classes.}
\label{fig:dfew_cm}
\end{figure}

Figure~\ref{fig:dfew_cm} shows the confusion matrix for the DFEW dataset. The system performs best on Happy expressions (84.9\% recall), with strong performance on Angry (68.4\%) and Sad (65.4\%). Notable confusion occurs between Neutral and Surprise (45 instances) and Neutral and Angry (42 instances), reflecting challenges in distinguishing subtle or ambiguous expressions in naturalistic in-the-wild video data, particularly when low-arousal states such as neutrality overlap with higher-arousal expressions.

\subsection{Ablation Studies}
\label{ssec:ablation}
Ablation studies are conducted on the DFEW dataset to systematically evaluate and justify the design choices of the proposed system.

\subsubsection{Embedding Dimension}
\begin{figure}[t]
\centering
\includegraphics[width=1\linewidth]{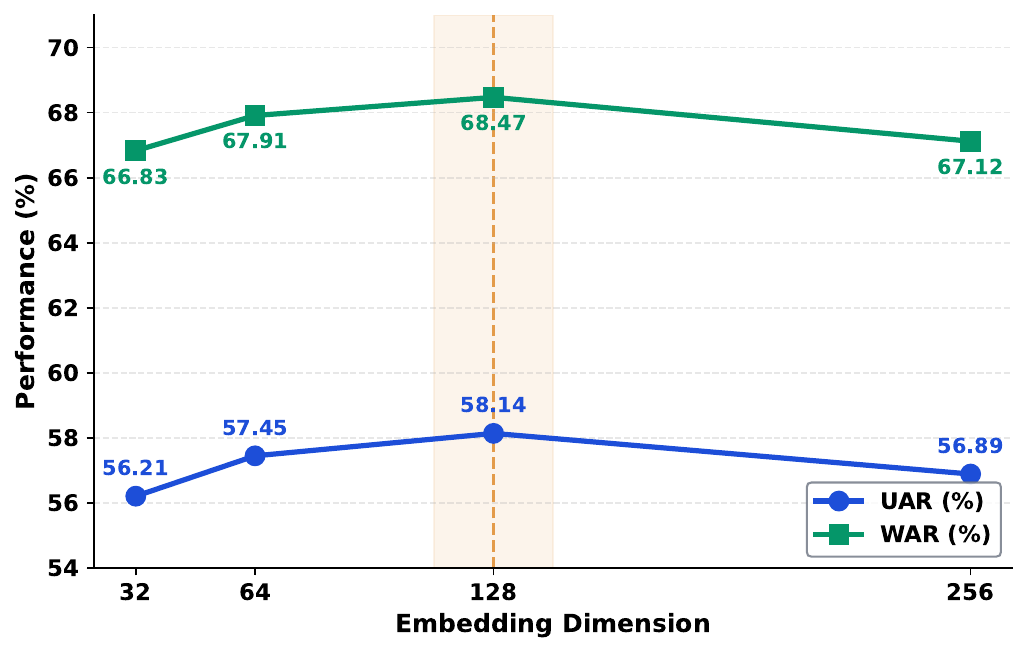}
\caption{Ablation study: effect of culture embedding dimensionality on DFEW.}
\label{fig:embedding_dim}
\end{figure}

The impact of the dimensionality of the culture embedding is investigated on the DFEW dataset. UAR and WAR with varying dimensionality are presented in Figure~\ref{fig:embedding_dim}. It may be noted that a 128-dimensional embedding achieves the best trade-off between representational capacity and generalization. Increasing the embedding dimension to 256 results in performance degradation, indicating that larger embeddings lead to overfitting without improving recognition performance. 

\subsubsection{Backbone Architecture Comparison}
The impact of the backbone architecture is evaluated by comparing a 3D CNN, TimeSformer \cite{bertasius2021space}, and ViViT under identical culture-agnostic training protocols on the DFEW dataset. Notably, ViViT achieves the highest performance with 53.14\% UAR and 62.71\% WAR, followed by TimeSformer (50.61\% UAR, 61.03\% WAR) and 3D CNN (48.34\% UAR, 58.79\% WAR). This validates the choice of ViViT as the backbone for the proposed system.

\subsubsection{Cultural Information Integration Strategies}
\begin{figure}[t]
\centering
\includegraphics[width=1\linewidth]{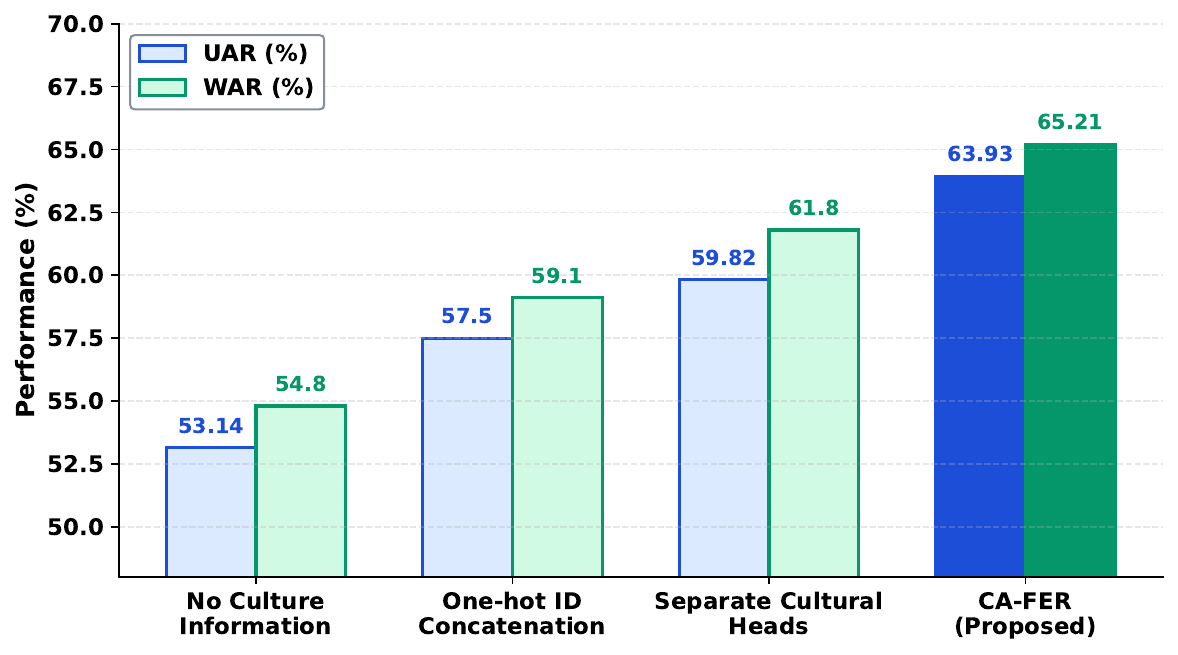}
\caption{Comparison of cultural information integration strategies on the DFEW dataset.}
\label{fig:conditioning_strategies}
\end{figure}

Different strategies for incorporating cultural information into facial representation learning are compared on the DFEW dataset, as shown in Figure~\ref{fig:conditioning_strategies}. The culture-agnostic baseline does not utilize any cultural information during representation learning. Introducing one-hot cultural identity concatenation improves performance by incorporating explicit cultural labels into the latent representation. Further improvement is observed using separate cultural classification heads, which enable culture-specific decision boundaries. The proposed CA-FER system achieves the best overall performance by adaptively recalibrating latent representations using behaviorally grounded cultural priors. Compared with the culture-agnostic baseline, one-hot concatenation, and separate cultural heads, the proposed system improves UAR by +10.79 pp, +6.43 pp, and +4.11 pp, respectively. These results demonstrate that behaviorally grounded cultural priors provide more effective cross-cultural adaptation than discrete cultural encoding strategies.

\subsubsection{Effect of Cultural Embedding Granularity}
\begin{figure}[t!]
\centering
\includegraphics[width=1\linewidth, trim=0 0.64cm 0 0, clip]{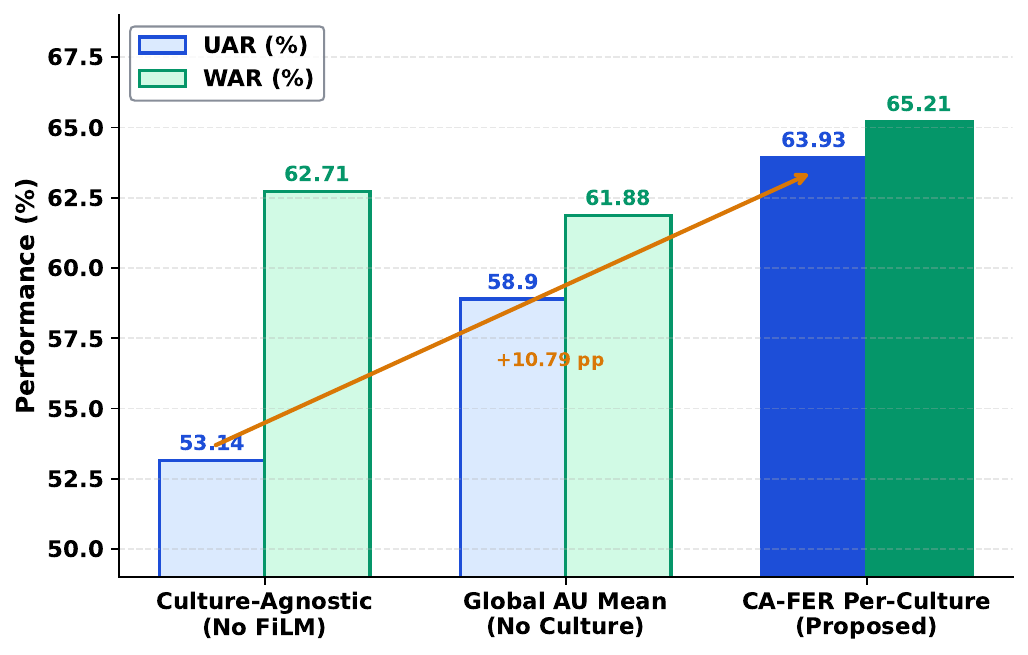}
\caption{Effect of cultural embedding granularity on the DFEW dataset.}
\label{fig:granularity}
\end{figure}

To isolate the contribution of culture-specific behaviorally grounded priors from the global mean cultural prior, an intermediate configuration is evaluated. In this setup, a single shared embedding is initialized from the global cultural prior across all cultures ($\mathbf{e}_\text{global} = \frac{1}{C}\sum_{c=1}^{C} \hat{\mu}_c$) and applied uniformly to all samples to recalibrate the latent features of the spatio-temporal model. As shown in Figure~\ref{fig:granularity}, this global-prior configuration achieves 58.90\% UAR, outperforming the culture-agnostic baseline by +5.76 pp. This demonstrates that the global prior provides a strong baseline improvement even without explicit cultural identity, as the global mean AU-based initialization captures shared facial behavior patterns across cultures. However, replacing this coarse global prior with culture-specific priors further improves UAR to 63.93\%, yielding an additional gain of +5.03 pp. This indicates that while global priors provide a useful behavioral foundation, culture-specific priors mitigate cultural bias much more effectively.

\subsection{Per-Culture Performance Analysis}
\label{ssec:culture_analysis}
Table~\ref{tab:culture_specific} presents a detailed performance breakdown across all four cultural groups in GCC-FER. Separate models were trained for each cultural group using 5-fold stratified cross-validation. Each model learned expression patterns specific to its respective culture.

\begin{table}[!htbp]
\caption{Culture-specific performance on the GCC-FER dataset.}
\label{tab:culture_specific}
\centering
\small
\begin{tabularx}{\columnwidth}{|X|c|c|c|}
\hline
\textit{Culture} & \textit{UAR (\%)} & \textit{WAR (\%)} & \textit{Samples} \\
\hline
Caucasian    & 63.1 & 62.7 & 1,159 \\
East Asian   & 60.6 & 62.9 & 912   \\
African      & 60.4 & 68.3 & 1,402 \\
South Asian  & 58.3 & 61.7 & 1,329 \\
\hline
\end{tabularx}
\end{table}

The Caucasian group achieves the highest UAR (63.1\%), indicating more balanced recognition across expression classes. The African group exhibits the largest UAR–WAR gap (7.9\%), indicating that the model performs well on dominant expression classes but struggles with underrepresented ones. This reflects the impact of class imbalance despite having the highest sample count. The South Asian group achieves the lowest UAR (58.3\%), reflecting the challenge of modeling subtle expression patterns specific to this cultural group. Furthermore, this distinctly motivates the need for culture-specific modeling approaches.

\subsection{Per-Expression Performance Analysis}
\label{ssec:emotion_analysis}
Further analysis for each facial expression is presented in Table~\ref{tab:per_emotion_combined} for both the GCC-FER and DFEW datasets. It may be observed that across both datasets, Happy achieves the highest F1 score (81.6\% on GCC-FER, 86.1\% on DFEW), consistent with existing DFER literature that identifies Happy as the most visually salient and culturally consistent expression~\cite{jiang2020dfew}. Disgust and Fear exhibit the most pronounced performance degradation, with Disgust achieving only 5.6\% recall on DFEW (18 samples) and 42.5\% recall on GCC-FER. This decline correlates directly with sample scarcity. Disgust constitutes approximately 3\% of the DFEW dataset, which is consistent with prior findings reporting 0.69--4.14\% recognition accuracy for Disgust~\cite{jiang2020dfew}. 

\begin{table*}[t]
\caption{Per-expression performance on GCC-FER and DFEW datasets}
\label{tab:per_emotion_combined}
\centering
\small
\setlength{\tabcolsep}{3pt}

\begin{tabularx}{\textwidth}{l*{8}{>{\centering\arraybackslash}X}}
\toprule
& \multicolumn{4}{c}{\textbf{GCC-FER}} & \multicolumn{4}{c}{\textbf{DFEW}} \\
\cmidrule(lr){2-5} \cmidrule(lr){6-9}
\textit{Expression} & \textit{Precision} & \textit{Recall} & \textit{F1} & \textit{Samples} 
                 & \textit{Precision} & \textit{Recall} & \textit{F1} & \textit{Samples} \\
\midrule

Angry    & 66.4\% & 50.9\% & 57.6\% & 690   & 59.1\% & 68.4\% & 63.4\% & 266 \\
Disgust  & 60.0\% & 42.5\% & 49.8\% & 254   & 33.3\% & 5.6\%  & 9.5\%  & 18  \\
Fear     & 62.5\% & 44.5\% & 52.0\% & 326   & 40.7\% & 35.8\% & 38.1\% & 123 \\
Happy    & 81.6\% & 82.3\% & 82.0\% & 1086  & 87.3\% & 84.9\% & 86.1\% & 325 \\
Neutral  & 56.9\% & 77.7\% & 65.7\% & 1189  & 58.7\% & 57.6\% & 58.1\% & 323 \\
Sad      & 68.0\% & 66.9\% & 67.4\% & 810   & 74.0\% & 65.4\% & 69.4\% & 266 \\
Surprise & 67.8\% & 51.6\% & 58.6\% & 432   & 48.0\% & 56.5\% & 51.9\% & 193 \\

\bottomrule
\end{tabularx}
\end{table*}
\subsection{Feature Space Visualization}
\label{ssec:tsne}
To qualitatively understand how cultural conditioning affects learned representations, t-distributed stochastic neighbor embedding (t-SNE) \cite{van2008visualizing} projections of 768-dimensional feature vectors are compared between the culture-agnostic baseline and the proposed CA-FER system. Features are extracted from the final layer preceding the classifier using GCC-FER validation samples from Fold 2 ($n \approx 4{,}000$). Both models use identical t-SNE parameters (perplexity = 30, 1000 iterations) on identical samples.

\begin{figure*}[t]
\centering
\includegraphics[width=\textwidth, trim=0cm 0cm 0cm 0cm, clip]{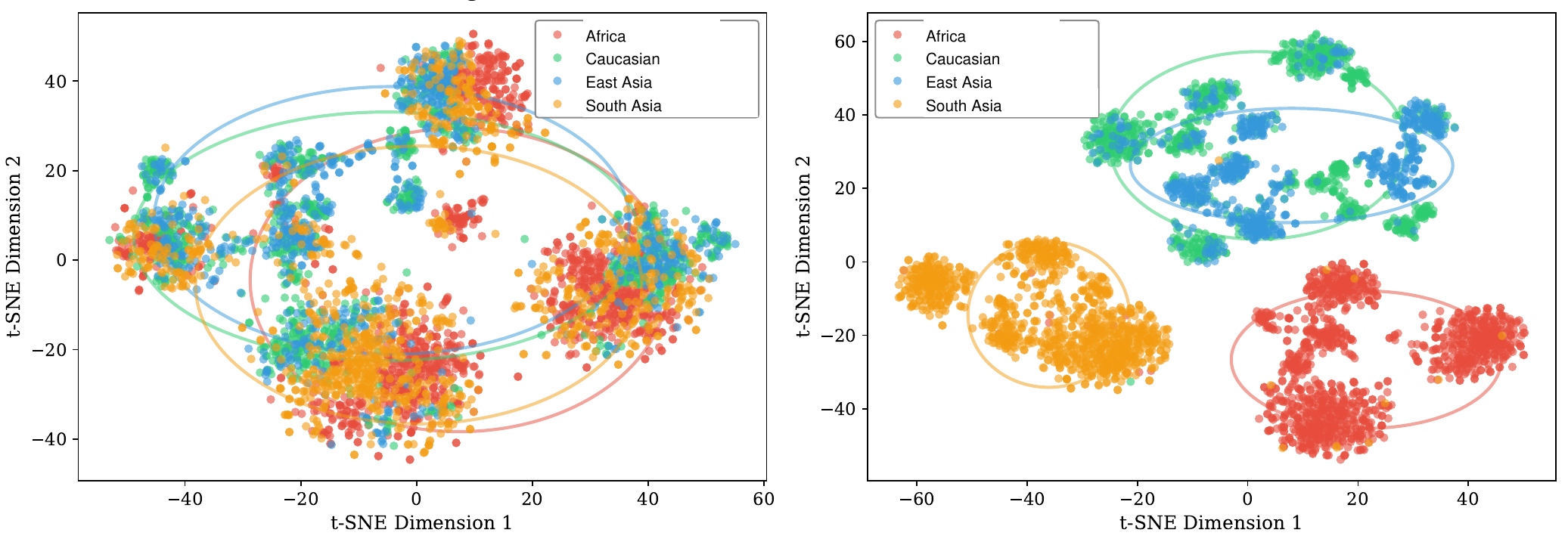}
\caption{t-SNE visualization of learned feature embeddings colored by cultural group. Left: The culture-agnostic baseline shows substantial inter-cultural overlap. Right: The proposed CA-FER system demonstrates clearer cultural clustering with reduced overlap (Africa: red, Caucasian: green, East Asia: blue, South Asia: yellow). Ellipses indicate approximate cluster spread.}
\label{fig:tsne_culture}
\end{figure*}

As presented in Figure~\ref{fig:tsne_culture}, the culture-agnostic baseline produces embeddings with substantial inter-cultural overlap, suggesting limited sensitivity to culture-specific facial behavioral patterns. In contrast, the proposed CA-FER system organizes the learned representation space into more distinct cultural regions while maintaining expression-discriminative structure. This indicates that the proposed cultural priors successfully capture culture-dependent variations in facial expression manifestations, thereby reducing cultural bias in the learned representation. The resulting feature organization provides an interpretable explanation for the per-culture performance improvements reported in Table~\ref{tab:culture_specific}.

\section{Conclusion and Future Work}
This work addresses culture-aware facial expression recognition in global context. A novel multi-cultural dynamic FER dataset, GCC-FER, is proposed. The GCC-FER dataset contains 23,934 video samples across four cultural groups and seven basic expressions, providing a diverse benchmark for evaluating multicultural FER systems. Unlike existing FER datasets that predominantly focus on culturally limited populations, GCC-FER includes underrepresented cultural groups, enabling the generation of behaviorally grounded cultural priors. By leveraging behaviorally grounded cultural priors, spatio-temporal latent features are adapted according to the culture. This provides a foundation system for mitigating cultural bias in facial expression recognition. Experimental results demonstrate that the proposed behavioral cultural priors consistently outperform culture-agnostic baselines. It is to be noted that the system achieves competitive performance on the DFEW benchmark and in-house GCC-FER dataset. Feature space visualization additionally indicates that cultural priors encourage the formation of structured, culture-sensitive representations, offering an interpretable explanation for the performance gains.

Overall, these findings highlight the importance of modeling cultural context in affective computing. Behaviorally grounded representations, particularly AU-based cultural embedding initialization, play a key role in improving performance across diverse populations. Finer-grained cultural hierarchies will be explored in future. Better modeling of minority expression classes will additionally be explored.

\section*{CRediT authorship contribution statement}

\textit{Sonalika Singh:} Investigation, Formal analysis, Writing - original draft, Data curation. 

\textit{Jyotirindra Dandapat:} Data curation, Writing - editing. 

\textit{Avishi Razdan:} Data curation. 

\textit{Kshipra V. Moghe:} Methodology, Writing - review \& editing. 

\textit{Puneet Gupta:} Supervision, Writing - review \& editing. 

\textit{Lalan Kumar:} Supervision, Writing - review \& editing, Project administration.

\bibliographystyle{IEEEtran}
\balance
\bibliography{refs}

\end{document}